%% file: sdnps-middleware2013.tex
\documentclass[runningheads,a4paper]{llncs}

\usepackage{amssymb}
\pdfoutput=1
\usepackage{graphicx}
\usepackage{makeidx}

\makeindex

\usepackage{url}

\newcommand{\publishsubscribe}{publish\slash subscribe}

\newcommand{\pubsub}{pub\slash sub}
\newcommand{\Pubsub}{Pub\slash Sub}

\begin{document}

\mainmatter  

\title{SDN-like: The Next Generation of \Pubsub\\ \small{\textit{Technical Report}}}

\titlerunning{SDN-like \Pubsub}

%
%
\author{Kaiwen Zhang \and Hans-Arno Jacobsen}
\authorrunning{}

\institute{University of Toronto}

%
%

\toctitle{Lecture Notes in Computer Science}
\tocauthor{Authors' Instructions}
\maketitle

\begin{abstract}
\input abstract.tex
\end{abstract}

\section{Introduction}
\input introduction.tex

\section{Background}
\input background.tex

\section{Centralized controller for \pubsub}
\label{sec:controller}
\input controller.tex

\section{Separation of control and data in \pubsub}
\input decoupling.tex

\section{Reference architecture}
\label{sec:architecture}
\input architecture.tex

\section{Use cases for SDN-like \pubsub}
\input usecases.tex

\section{Conclusion}
\input conclusion.tex

\bibliographystyle{splncs03}
\bibliography{sdnps-middleware2013}

\appendix

\end{document}

%% file: abstract.tex
Software-Defined Networking (SDN) has raised the boundaries of cloud
computing by offering unparalleled levels of control and flexibility
to system administrators over their virtualized environments. To
properly embrace this new era of SDN-driven network architectures, the
research community must not only consider the impact of SDN over the
protocol stack, but also on its overlying networked applications. In
this big ideas paper, we study the impact of SDN on the design of
future message-oriented middleware, specifically \pubsub\ systems. We
argue that key concepts put forth by SDN can be applied in a
meaningful fashion to the next generation of pub/sub
systems. First, \pubsub\ can adopt a logically centralized controller
model for maintenance, monitoring, and control of the overlay
network. We establish a parallel with existing work on
centralized \pubsub\ routing and discuss how the logically centralized
controller model can be implemented in a distributed manner. Second,
we investigate the separation of the control and data plane, which is
integral to SDN, which can be adopted to raise the level of decoupling
in \pubsub. We introduce a new model of \pubsub\ which separates the
traditional publisher and subscriber roles into flow regulators and
producer/consumers of data. We then present use cases that benefit
from this approach and study the impact of decoupling for performance.

%% file: introduction.tex
Software-Defined Networking~\cite{sdn} offer unprecedented levels of
flexibility and control to administrators of cloud-based
datacenters. By employing a logically centralized controller, which
raises the level of abstraction of networking components into a single
unified view, the complexity of configuring the network is
significantly reduced, which translates to cost savings for large
corporate datacenters. Furthermore, the separation of the control and
data planes allows for more flexible and powerful routing policies to
be expressed, which results in improved performance. As an example,
Google's inter-datacenter WAN powered by an SDN has achieved 95\%
network utilization~\cite{googlesdn}. While still in an early stage,
SDN is expected to fully realize the vision of the
Infrastructure-as-a-Service provider model.

In light of this new approach to networking, the research community
must adapt networked applications to harness the potential of SDN. One
fundamental type of applications is message-oriented middleware, such
as \pubsub~\cite{manyfaces}. Since those applications are abstractions
of the underlying network layer, it is essential that they are capable
of leveraging the capabilities of the protocol stack. For
instance, \cite{sdnps} proposes a \pubsub\ architecture where the
controller is employed to disseminate routing information to
SDN-enabled switches using the content-based matching semantics of
the \pubsub\ system. By mapping \pubsub\ content to OpenFlow flow
entries, the resulting proof-of-concept system is capable of achieving
line-speed matching and dissemination of publications.

%
%

The above line of work represents the most straightforward approach to
integrating SDN with \pubsub. However, we argue that there exists a
more fundamental study of SDN which will benefit \pubsub. By
distilling the key concepts behind the success of SDN, we can reason
about their relative impact within the context of \pubsub. At a higher
level, many of the concerns encountered in the networking community
have an equivalent in \pubsub. For instance, overlay construction
for \pubsub\ is analogous to topology construction of the underlying
network. Along the same lines, we can extrapolate that the design
principles behind SDN, which are motivated by traditional networking
issues, can be applied to \pubsub\ to solve similar challenges.

In this big ideas paper, we present a new model of \pubsub\ which is
``SDN-like''. We build our model based on two key concepts of SDN: The
logically centralized controller and the separation of the control and
data plane. Enhancing the traditional \pubsub\ architecture with a
centralized controller raises the level of control over the topology
of the network by introducing a monitoring entity and a bootstrapping
process. The separation of the control and data plane can be achieved
in \pubsub\ through the decoupling of the traditional publisher and
subscriber roles into advertiser/producer and interest
manager/consumer, respectively. We describe how separating the control
plane (advertisements and subscriptions) and the data plane
(publication production and consumption) in \pubsub\ pushes further
onward the commonly established idealogies of \pubsub\ and can benefit
certain application scenarios. This decoupling allows for more
powerful features to be supported by \pubsub\ systems, such as
installing specific control policies over a subset of publishers and
subscribers.

The contributions of this paper are thus as follows:
\begin{itemize}
\item We propose a new breed of SDN-like \pubsub, where we argue 
that integrating key concepts borrowed from SDN can solve analogous
problems in \pubsub.
\item We present the design of a logically centralized controller for 
monitoring and controlling the \pubsub\ overlay.
\item We advocate and demonstrate the benefits bedind a division of existing 
\pubsub\ roles to decouple the control and data  plane: Advertiser/producer 
for publisher and interest manager/consumer for subscriber.
\item We illustrate that the support for policy-based control of \pubsub\ 
agents will allow for more meaningful access to data.
\item We present a proof-of-concept architecture with the 
matching protocol suite to power our SDN-like \pubsub\ model.
\item Finally, we present a study of use cases that will benefit from our model,
\end{itemize}

%% file: background.tex
In this section, we describe in more detail our \pubsub\ reference
model. We also elaborate on SDN, its key concepts, and the advantages
it offers over traditional networking.

\subsection{Content-based \publishsubscribe}
\label{sec:psbg}

\Pubsub\ is a messaging paradigm that allows for loosely coupled communication 
between data producers (called publishers) to data consumers (called
subscribers). Events (referred to as publications) flow from the
publishers through an overlay network of brokers, which route the
message traffic towards the intended recipients. In the content-based
matching model~\cite{siena}, subscribers can express their interest as
a conjunctive list of predicates. Publications, which are lists of
attribute-value pairs, are considered to match if they satisfy all the
predicates of a subscription, in which case they must be delivered to
the subscription source.

In terms of routing, a variety of routing algorithms
exist. Advertisement-based and subscription-based forwarding
respectively flood advertisements or subscriptions through the
overlay~\cite{content-based}. In the case of advertisement-based
forwarding, the subscriptions are floated towards matching publishers
using the reverse path from the advertisements. The delivery tree for
publications is computed on a hop-by-hop basis, with each broker
matching the publication to the next set of hops containing a matching
subscription.

Another routing model we consider is \emph{rendezvous-based}. One
broker in the overlay is designated as the rendezvous node and is in
charge of computing the \pubsub\ matching~\cite{hermes}. The rest of
the brokers simply relay all incoming traffic to the rendezvous
node. This model simplifies the routing process and reduces the
matching overhead by having a centralized matcher, at the expense of
non-optimal routing paths and potentially limited scalability.

We classify the publication flow as constituting the \emph{data
plane}, whereas the control plane is comprised of
advertisements and subscriptions. The \pubsub\ system is therefore in
charge of disseminating the message traffic in the data plane to the
correct recipients as guided via the information from the control
plane.

One key property of \pubsub\ is its loosely coupled nature. This
decoupling typically spans three dimensions: Space, time and
synchronization. By employing a \pubsub\ middleware, data producers
and consumers are not aware of the identities of other participants,
at the time the data is produced or consumed, or they need not to
directly exchange data with one another. Decoupling is a critical
aspect of \pubsub, since the lack of coordination allows \pubsub\ to
leverage highly asynchronous protocols for scalability. Furthermore,
decoupling reduces the complexity of the \pubsub\ model, which
supports only a limited number of functions. This decoupling is made
possible due to the variety of application scenarios with loosely
synchronized distributed components for which \pubsub\ constitutes an
ideal choice as a dissemination layer.

\subsection{Software-defined networking}

SDN is a new paradigm for networking, abstracting the underlying
network as an unified entity, which can be manipulated through a
controller~\cite{sdn}. Although not a clean-slate proposal for a
future Internet architecture, SDN does require specialized hardware
for switches and routers to be controllable from a remote location
(i.e., controller). The two properties of SDN we focus on are:

\textbf{Separation of the control and data plane}: SDN-enabled switches receive
routing instructions in the form of policies over the traffic
metadata. This is realized through the OpenFlow standard, which
classifies traffic into multiple flows, with each flow associated to a
sequence of actions~\cite{openflow}. For instance, a simple
OpenFlow-enabled \pubsub\ solution could classify flows per topic,
with each topic being associated to a multicast group corresponding to
the subscribers of that topic.

\textbf{Logically centralized controller}: The aforementionned decoupling 
is leveraged through the use of a logically centralized controller,
such as NOX~\cite{nox} or Floodlight~\cite{floodlight}. The controller
maintains a unified view of the network and enforces a set
of global policies. Common applications for a controller include
monitoring of the topology and dynamic rerouting for load-balancing
purposes. Although logically centralized, the controller itself can be
scaled through physical distribution~\cite{hyperflow}.

Both properties together significantly reduce the complexity of
configuring individual switches, since they only respond to a single
SDN controller. Furthermore, the centralized view of the network
allows for more dynamic policies to be enforced by the controller,
resulting in more efficient and scalable networks.

%% file: controller.tex
We propose the use of a logically centralized controller in \pubsub\
for monitoring and reconfiguration of the overlay broker network. The
unified view of the overlay allows for dynamic routing policies in
case of congestion and failures, akin to its counterpart in SDN. In
addition, we employ the \pubsub\ controller during a boostrap process,
to setup the overlay and disseminate configuration properties to the
brokers and clients.

\begin{figure}[t]
\centering
\includegraphics[clip=true,width=\linewidth]{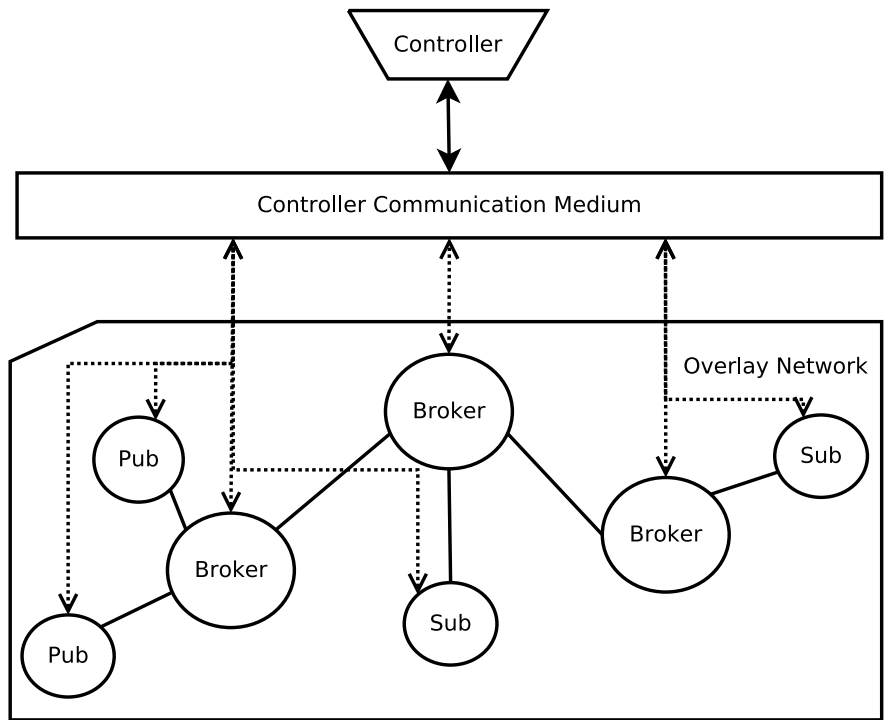}
\caption{Controller architecture}
\label{fig:architecture}
\end{figure}

Fig.~\ref{fig:architecture} shows the general architecture of
a \pubsub\ system with a controller. Every node in the system, which
includes clients (publishers and subscribers), must establish a
connection to the controller through a bootstrapping process. The
controller can return the necessary configuration properties to the
node. For instance, new brokers require the identity of neighbors
within the determined topology. The bootstrapping process could also
be used to provide a unique identifier to each node.

Communication to the controller is regulated through a separate
channel as it is the case in SDN. The separation of channels reduces
the complexity associated with handling mixed traffic over one channel
and allows this channel to be tuned for communication with the
controller. For instance, a client needs to establish only a
short-lived TCP connection to the controller for bootstrapping
purposes. Passing this traffic through the overlay would unnecessarily
burden the system.

The main role of the logically centralized controller is topology
maintenance. By having global knowledge of the network, the controller
can adaptively reconfigure the topology to be fault-resilient and
efficient (i.e., exhibit low node fanout, be forced acyclic, or have a
set number of redudant path between every pair of system edge brokers,
where redundancy may go as far as requiring the existance of
physically disjoint path.) Although there exists a number of
decentralized solutions for efficient overlay
construction~\cite{chendc}, we argue that providing global knowledge
can help the controller make more informed decisions. This design
would eschew the need for complex distributed protocols, which reduces
the complexity of operating brokers. In this setting, the brokers are
lightweight entities which listen to instructions from the logic
maintained at the controller. Furthermore, the controller requires an
incremental overlay construction algorithm since it is aware of churn
in the system. Currently incremented solutions for overlay
constructions are reactive in nature~\cite{chentr} and seek to
restablish sought properties after the churn has occured. By enhancing
the controller with global knowledge, we should be able to achieve a
higher prediction accuracy and create topologies that are resilient
despite churn.

Another role for the centralized controller is dynamic routing
policies. We can establish a parallel between the publication space
of \pubsub\ and the flows of OpenFlow to maintain on-demand quality of
service. For instance, publications of a certain kind (i.e., a certain
topic or including certain attribute-value pairs) could be
prioritized. To solve congestion, traffic can be labelled according
to \pubsub\ meta-data and rerouted to alternative paths. To support
this feature, the logically centralized controller needs to fetch data
from brokers on a regular basis to monitor the current state of the
traffic. This requires the brokers to maintain long-lived connections
to the controller, which raises scalability questions. To address this
problem, we can either employ another overlay-based channel for
broker-controller interactions (such as another \pubsub\ system), or
distribute the controller (while maintaining logical centrality).

%% file: decoupling.tex
%
%
In Sec.~\ref{sec:psbg}, we described how \pubsub\ decoupling enables
publishers and subscribers to produce and consume data without space,
time or synchronication coupling. The lack of coordination and the
limited degree of coupling gives the \pubsub\ system ample flexibility
to provide relaxed consistency and reliability guarantees while
maintaining simple matching semantics in a scalable and efficient
manner. However, one dimension which remains coupled is the control
and data plane. In other words, the producers and consumers of data
in \pubsub\ remain in control of the data itself. For instance, this
means that a subscriber submits subscriptions in order for content to
be delivered to itself. In this case, we see that the subscriber
controls the type of data that is being delivered to itself. By
explicitly sending its own subscriptions, each subscriber is
completely aware of the type of publications it expects to receive.

We seek to lift this binding by introducing the concept of control and
data decoupling in \pubsub. To do so, we break down the traditional
roles of publishers and subscribers into advertisers and producers as
well as interest managers and consumers, respectively. Advertisers and
interest managers operate on the control plane, installing policies
which regulate the advertisements (ads) and subscriptions (subs) of
producers and consumers, respectively. At the most fundamental level,
advertisers and interest managers are able to use \emph{remote
advertisements and subscriptions}, i.e., perform advertising and
subscribing on behalf of other clients. We thus broke down the
coupling of the data and control flows by allowing for controls to
direct data to a different destination as opposed to from where the
controls were originating.

%
%

We argue that the data and control binding in traditional \pubsub\ is
an an artificial construct with no real justification behind it. Due
to the loosely coupled nature of \pubsub\ applications, there was no
need to differentiate between data and control. The use cases assume
that clients either formulate simple ads or subs over a set of
predetermined topics (channel-based) or a limited publication space
(topic-based or content-based). Since there is no coordination amongst
clients once the system is online, publishers and subscribers are
given the freedom to control their message content. However, we have
discovered that new applications for \pubsub\ contain complex
specifications which require more sophisticated \pubsub\ semantics,
which can translate in uncertainties in the advertisement and
subscription spaces~\cite{liu04}. In other cases, \pubsub\ is employed
to support Big Data applications where subscribers are not expected to
consume the entire stream of publications, but would rather limit
themselves to a summary, which could be obtained via top-k
filtering~\cite{zhangtopk} or aggregation~\cite{asiaPosterMW2012}.

In those situations, the data sinks and sources might not be the most
suitable entities to formulate their own ads or subs. A third party,
equipped with more intimate knowledge of the current state of the
system, is able to express or transform ads and subs
accordingly. Another possibility is for the third party controller to
leverage the flexibility in ads and subs to enhance the internal
performance of the \pubsub\ system in a transparent manner to the
end-user. Separating the control and the data plane allows for remote
parties to make such adjustments whenever adequate, for example, when
the principal entities would be unable to do so.

\begin{figure}[t]
\centering
\includegraphics[clip=true,width=\linewidth]{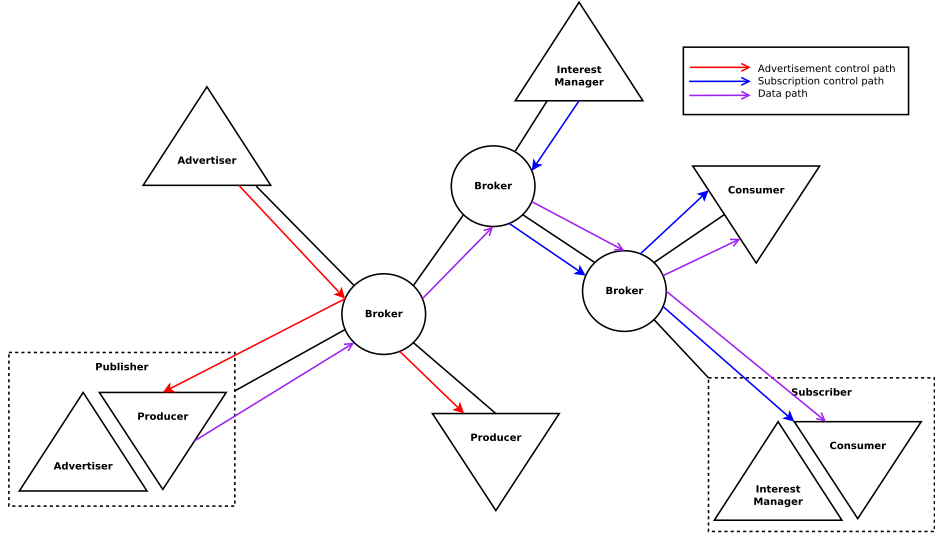}
\caption{Mixed SDN-like topology with decoupled control and data plane}
\label{fig:decoupling}
\end{figure}

Furthermore, the model we propose can be considered a generalization
of \pubsub. This decoupling is orthogonal to other properties
of \pubsub\ and to its matching semantics. The traditional roles of
publishers and subscribers are retained as physically co-located data
and control entities. From the end-user point of view, each pair of
roles can be considered as a logically unified publisher or
subscriber. However, the SDN-like \pubsub\ system is agnostic to the
binding and considers each role in the pair as separate entities. What
the decoupling allows is for data consumers to receive data without
specifying its own subscriptions or for a subscriber, who has its own
subscriptions, to have its subscription space altered by remote
interest managers. We therefore envision a model where mixed roles are
compatible with one another: The system can allow controllers, data
sources and sinks, to be located anywhere in the topology (see
Fig.~\ref{fig:decoupling}). Advertisers and interest managers are able
to control the data flow from and to publishers and subscribers as
well as producers and consumers alike.

We also note that while the separation of concerns is inspired from
SDN, the purpose of the decoupling is different. In SDN, the
separation works in tandem with the centralized controller whereas in
SDN-like, the logically centralized controller presented in
Sec.~\ref{sec:controller} is by and large orthogonal to this
property. While it is feasible for the centralized controller to act
as a producer and interest manager, it is not a necessary
condition. For instance, we allow multiple advertisers to effect the
same producer. The relationship between control roles can be
structured either as a hierarchy, where higher level entities have the
ability to modify or override the ads and subs of lower level ones, or
as a collection of administrative domains, where each control entity
is in charge of regulating a region of the publication
space. Implementation of SDN-like \pubsub\ will however require
changes to the routing protocols, which we will discuss in
Sec.~\ref{sec:architecture}.

We now delve into the details of the new roles, their principal usage,
and the new functions required to empower those roles.

\subsection{Advertiser and producer roles}

We define the functions of the advertiser at two levels:
advertisements and publications.

\subsubsection{Advertisement level -} 

First, advertisers are allowed to create \emph{advertisement policies}
which regulate what advertisements a producer will obtain. In a
traditional \pubsub\ model, an advertisement is an auxiliary type of
control message used to set routing paths for a publisher. In our
context, we leverage the property that a publisher must advertise over
a certain space before publishing within that region to employ
advertisements as an \emph{access control} mechanism.

Since producers are decoupled from advertisers, they are not aware of
the advertisements they possess and can potentially publish data
outside of their advertised range. In that case, two scenarios can
occur: Either the publication is silently dropped, or a feedback
message is returned to the source. Both type of responses are valid
depending on the application. Silently dropping publications might be
useful for lightweight publishers (e.g., sensors) that are specialized
in outputting events at a fixed frequency, while feedback might guide
a self-tuning publisher to internally filter on its own before
publishing, or to package its publications in a format that is
compatible with the advertisements.

Beside the traditional advertisements targeting single producers,
advertisers can enact advertisement policies based on the metadata
supplied by producers. One possibility is to employ policies when
producers publish data in a highly structured manner. For example,
each publisher advertises on the topic of its own country. This can be
formulated as an advertisement policy which extracts the country from
each advertiser's metadata and produces the corresponding
advertisement. Although this type of simple advertisements could
easily be advertised manually by each publisher using existing models,
the presence of an advertisement policy creates an invariant which can
be leveraged to optimize the system, either through clustering
techniques or by pruning the data structure storing advertisements.

Another way to employ policies is by altering existing
advertisements. A monitoring advertiser can prioritize certain
producers by restricting others and reducing their advertising
space. Top-k filtering can also be achieved at a per-source level by
analyzing publications from various producers and ``turning off''
those less relevant.

\subsubsection{Publication level -}

Secondly, advertisers can be equipped to express \emph{publication}
policies. Those policies can formulate the appropriate publication
semantics to attach to a particular content to be published. For
example, using producer metadata and by inspection of a publication
payload, a broker can decide to publish that publication over a certain
topic. This strategy could be used for top-k filtering: A producer
could start off with its publications being disseminated in a lower
level channel before the advertiser learns enough about the producer
to elevate it to a higher status, and delivering its future
publications into a topic for higher quality content.

Another reason for using \emph{publication} policies is to push the
decoupling of \pubsub\ semantics even further. By
using \emph{publication} policies, the producer does not need to be
aware of \pubsub\ semantics at all. It is simply handing this data to
the broker, which attaches to it the necessary \pubsub\ header such
that the content will be forwarded to interested parties. For security
and privacy issues, it might be desirable to expose as little
information as possible about the \pubsub\ system to the producers so
they cannot infer anything about the system based on the \pubsub\
semantics they publish on.

\subsection{Interest manager and consumer roles}

At the receiving end, the decoupling of control and data can be used
to satisfy very dynamic subscription patterns. Interest managers
submit \emph{subscription policies} which control the subscriptions
consumers have based on collected metadata and the state of their
current subscriptions.

One use for subscription policies is to support fine-grained
subscriptions with constant churn. Due to a massive publication space,
certain applications demand that their subscribing clients submit very
fine-grained subscriptions and continuously increment them as
needed. This is, for instance, true for location-based applications,
where subscribers subscribe to a small area around their current
location and continuously update as they move. In such cases, a
subscription policy can be parametric (see~\cite{parametric}) to serve
as a template for a subscription everytime the location metadata is
updated. As for advertisements, the policy serves as an invariant
which can be taken into account when optimizing the system.

Another use is for consumers with fuzzy or unknown interest. In such
cases, the consumer may be lacking the necessary knowledge about the
state of the system to create the precise subscription. In the
traditional model, the subscriber would have to subscribe first to a
larger space of content and then downsize it to the relevant content,
which is inefficient. Another possibility is that the interest of a
consumer is conditional upon the state of the system. For instance, it
is conceivable for a consumer to subscribe to the publications
belonging to topics associated with publishers that are located near
the subscriber. In order for a subscriber to create those
subscriptions, it would have to subscribe to the position updates of
every publisher and create new subscriptions whenever one of those
publishers is within range of the subscribe. A much better approach is
to let an interest manager keep track of the position of every
producer and consumer in the system and remotely generate
subscriptions whenever a consumer is close to a producer. In such
scenario, it is sufficient for a single interest manager to monitor
the position updates of all entities in the system, rather than having
every subscriber monitor everything.

Finally, subscription policies can be used for generating
subscriptions for flexible consumers that are efficient with regards
to the current state of the system. This is the case when subscribers
only require a summary or a sample of the most relevant
publications. The interest manager can gauge the granularity of the
subscription to satisfy such consumers. Another possibility is that a
consumer is currently interested in multiple topics but do not require
data from every topic. A smart interest manager can evaluate, based on
current traffic amongst other things, which subscriptions particular
consumers can unsubscribe to alleviate the load of the system.

%% file: architecture.tex
We present a reference architecture for SDN-like \pubsub. This will server as a proof of concept by demonstrating the feasibility of our model, shown through the reusability of existing \pubsub\ components (such as the matching engine) to serve the additional functionalities required.
We also explain the various protocols and interactions found in our model.

The architecture and its subscription language is based on PADRES~\cite{fidler05}, a content-based \pubsub\ system operating over a federated overlay broker network. Publishers and subscribers are considered clients - we will extend that notation for advertisers/producers and interest managers/consumers. Each client connects to a single broker to access the system (called edge broker). The overlay consists of brokers connected to one another, which each broker only aware of its neighbors. The topology can be cyclic: the routing protocol ensures cycle-free delivery paths~\cite{cyclic}.

\subsection{Controller architecture}

Fig.~\ref{fig:controllerarch} lists the various components of the centralized SDN-like controller:

\begin{figure}[t]
\centering
\includegraphics[clip=true,width=\linewidth]{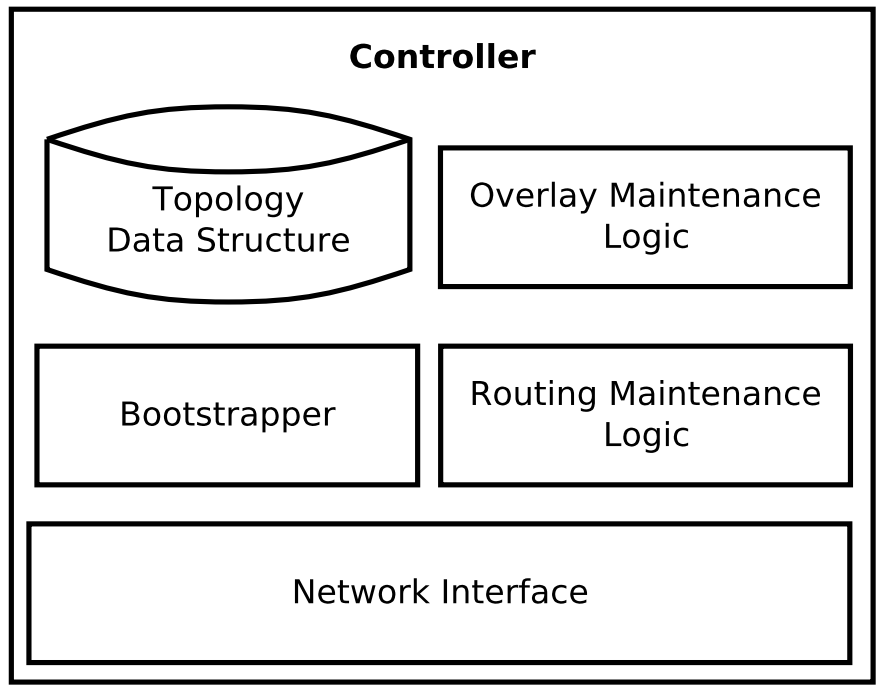}
\caption{Controller components}
\label{fig:controllerarch}
\end{figure}

\begin{description}
\item[Bootstrapper] \hfill \\
Each node joins the system by first contacting the controller. The bootstrapper module builds the configuration properties to be installed by the node. A counter is maintained by the bootstrapper to assign an unique identifier to the node. The bootstrapper also polls the overlay maintenance component to determine the placement of the node in the overlay. If it's a client, the bootstrapper will determine which edge broker it must connect to. For a broker, it builds a list of neighboring brokers to connect to. The topology data structure is then updated.
\item[Network interface] \hfill \\
Nodes establish a direct connection to the controller via TCP. Brokers must maintain long-lived connections and listen for instructinos from the controller, while clients can disconnect once they pass the bootstrapping sequence. Clients may receive commands from the controller via their edge broker. Clients must also establish a connection to the controller when leaving the system, which will remove the node from local storage.
\item[Topology data structure] \hfill \\
The controller collects various statistics about the brokers, which are packaged as regular broker information messages. These messages also serve as heartbeat messages for failure detection purposes. The information is compiled in the topology data structure, where the statistics for individual brokers is maintained.
\item[Overlay maintenance logic] \hfill \\
This component runs an incremental overlay construction algorithm to decide the optimal placement of a node at the moment it enters the system. Furthermore, the overlay is maintained by regularly monitoring the state of the topology as stored in the controller, and computing the necessary overlay transformations required. Those operations are then sent to the appropriate nodes. In order to contact a client, the command is instead sent to its edge broker. For instance, a broker may be asked to shed clients to a different broker. Those clients will then proceed in a transitionary migration phase~\cite{mobility}. The overlay maintenance is also in charge of repairing the topology when failures are detected. Clients who were previously connected to the failed node will eventually time out and restart the bootstrapping process in order to be assigned to a new edge broker.
\item[Routing maintenance logic] \hfill \\
The routing maintenance module leverages the information stored to detect congested nodes and links. Using a cost model, the component evaluates possible routing changes to alleviate congestion. If none are satisfactory, the component instead looks for possible topology reconfiguration. Any changes made are stored in the topology database and communicated to the overlay maintenance logic.   
\end{description}

\subsection{Policy mechanisms}

We now describe the mechanisms associated with policies. This involves the definition of metadata and policy languages, routing the related information and generating new advertisements and subscriptions.

\subsubsection{Metadata language}

In order to interact with policies set by advertisers/interest managers, producers and consumers expose some metadata to the \pubsub\ system. Metadata contains basic information about the client (such as ID) as well as application-specific information (such as geographical location). 
Metadata messages are formulated in the same format as publications, as lists of attribute-value pairs.

A producer/consumer is responsible for updating its metadata whenever a change occurs by sending a new metadata message. A tradeoff between the size of the metadata and the update frequency is therefore observed. Including highly dynamic data in the metadata will trigger a high update rate which can cause a significant overhead for the system.

\subsubsection{Policy language}

Policies contain two parts. The first part is a list of predicates which are matched against the state of every producer/consumer, which consists of its metadata and current advertisements/subscriptions. The second part is the set of instructions to be followed whenever a matching producer/consumer is found.

Essentially, the conditional section of a policy takes on the same format as a subscription. We can therefore reuse the matching engine to match policies over metadata. The difference is that policy matching constitutes the reverse operation: instead of matching a publication against known subscriptions, we are instead matching subscriptions (\emph{policies}) against known publications (\emph{metadata}). Since publications can be considered subscriptions containing solely equality predicates (ie. no ranges), data structures used to store subscriptions can easily be adapted to store metadata. What is needed is an indexing structure to query subscriptions and advertisements by source for fast lookup during policy matching involving the advertisement/subscription status of a given producer/consumer.

Once matched, the possible instructions include:
\begin{description}
\item[Insert ad/sub x] Create the specified advertisement/subscription for the matched target. The ad/sub can contained variable data which is replaced by metadata fields at the time of generation.
\item[Insert unad/unsub x] Finds ads/subs matching the specified argument x and send an unadvertisement/unsubscription message to remove x. If x is not found, ignore this command.
\item[Modify x y] Finds ads/subs matching x and modify them according to y. y contains modifications to the ad/sub, which could be in the form of insertion, deletion, or modifications of predicates. y can use read values for attributes in the existing predicates of x.
\end{description}

\subsubsection{Routing}

Metadata and policies can routed according to a variety of strategies. Strictly speaking, they must be routed to a quorum of brokers in order to guarantee that at least one can match every pair of metadata and policy. We propose three different strategies:
\begin{description}
\item[Metadata flooding] Metadata are flooded in the network. Policies can then be matched directly at the edge broker of corresponding advertiser/interest manager. This strategy is beneficial only if the workload contains more metadata than policies. The disadvantage is that edge brokers connected to advertisers/interest managers must collect all advertisements (already done in advertisement-based forwarding) and all subscriptions in order to compute matches involving ads/subs.
\item[Policy flooding] Policies are flooded in the network. Metadata can then be matched directly at the edgde broker of the producer/consumer. This approach also has the advantage that it is compatible with any forwarding model, since the edge broker will contain all the subscriptions of collected metadata and therefore has the necessary knowledge to computing matches for those producers/consumers.
\item[Rendezvous-based] Policies and metadata are routed towards a designed rendezvous broker. This broker also require complete knowledge of all ads and subs and can therefore compute matches for every producer/consumer. This approach is beneficial if used in conjunction with the rendezvous based forwarding approach for ads/subs.
\end{description}

%% file: usecases.tex
We now present a sample of application scenarios that motivate our design. Most of the use cases presented are already established in the \pubsub\ literature: we demonstrate how our design facilitate or enhance such applications. We also present novel uses of \pubsub\ possible with our model.

\subsection{Big Data analytics}

Support for event-based solutions have been rising in the area of Big Data analytics. Complex event processing systems are cited as the enabler for realtime analytics, with industrial support from companies such as Twitter (Storm~\cite{storm}). Metrics data collected at real-time are considered first citizens and processed to manipulate business processes. For instance, \pubsub\ can be used for dynamic service composition in service-oriented architecture~\cite{soa}, with each service provider modelling their specifications and interactions as subscriptions and publications respectively. In certain cases however, the business logic spans across services and requires the use of coordinators to enforce global safety constraints~\cite{yoon:soa}. In this case, the function of these coordinators is provided by our control nodes.

\subsection{Application performance management}

Another example can be found in the APM use case. Here, metrics data is collected to monitor the health of the system. For instance, an intrusion detection system operates by filtering over a stream of incoming events and dispatching alerts over anomalies. This can be supported using the content-based \pubsub\ model to obtain the precise substream of outlying events. However, the monitoring subscription might be frequently changing as it tunes itself to statistical values of the metrics attributes, which requires parametric subscriptions~\cite{parametric}. Such type of subscriptions are easily accomodated by our interest manager abstraction, which can fetch the required values (eg. via a key-value store) and install the appropriate the subscriptions at runtime.

\subsection{Client notification for long-lived objects}

Publish/subscribe can be used to notify or disseminate updates about a particular object. This is the case for Google, which developed its own \pubsub\-based notification system for its web services~\cite{thialfi}. Another example includes massively multiplayer online games, where players obtain a partial view of the game shown through its respective client. \Pubsub\ subscriptions are used to model the interest of clients vis-\`{a}-vis objects found in the game~\cite{boulanger06}. Players can subscribe to game objects based on location (eg. nearby objects), social relationship (eg. friends), and other attributes (eg. ranking of the player). The amount of game state a player needs to formulate accurate subscriptions can be prohibitely large that it can defeat the purpose of employing interest management techniques in the first place. To alleviate this problem, we can employ our interest managers to collect larger amount of data about the game and compute precise subscriptions for the players, thus leveraging interest management techniques more efficiently.

%% file: conclusion.tex
We presented \emph{SDN-like}, a new \pubsub\ model which borrows properties from software-defined networking. Our design for a centralized controller is used for bootstrapping purposes and topology control. We then extended \pubsub\ decoupling along the control/data plane and presented its benefits through example use cases. Our reference architecture shows how SDN-like can be implemented in a modular manner on top of an existing \pubsub\ engine.

While our work is inspired by SDN, the model itself can be realized without it. However, our future work will seek to implement to use an SDN-enabled underlying network (eg. using OpenFlow) to drive our SDN-like design in an efficient manner. 